\documentclass[11pt,british,english]{article}
\usepackage{mathptmx}
\usepackage[T1]{fontenc}
\usepackage[latin1]{inputenc}
\usepackage[pdftex,letterpaper]{geometry}
\usepackage{amstext}
\usepackage{graphicx}
\usepackage{epstopdf}

\makeatletter
\newcommand{\lyxaddress}[1]{
\par {\raggedright #1
\vspace{1.4em}
\noindent\par}
}

\usepackage{times}

\usepackage{geometry}

\usepackage{setspace}



\makeatother

\usepackage{babel}

\begin{document}

\title{\textbf{Theoretical signal-to-noise ratio of a slotted surface coil
for magnetic resonance imaging}}

\maketitle
\vspace{10mm}

\begin{center}
K. Ocegueda, S. S. Hidalgo, S. E. Solis, A. O. Rodriguez%
\footnote{\noindent \begin{center}
Corresponding author: Alfredo O. Rodriguez, email: arog@xanum.uam.mx.
\par\end{center}%
}$ $
\par\end{center}

\noindent \begin{center}
Departamento de Ingenieria Electrica, Universidad Autonoma Metropolitana
Iztapalapa, Av. San Rafael Atlixco 186, Mexico D. F. 09340. Mexico. 
\par\end{center}

\vspace{10mm}

\begin{abstract}
The analytical expression for the signal-to-noise ratio of a slotted
surface coil with an arbitrary number of slots was derived using the
quasi-static approach. This surface coil based on the vane-type magnetron
tube. To study the coil perfomance,\emph{ }the\emph{ }theoretical\emph{
}signal-to-noise ratio predictions of this coil design were computed
using a different number of slots. Results were also compared with
theoretical results obtained for a circular coil with similar dimensions.
It can be appreciated that slotted surface coil performance improves
as the number of coils increases and, outperformed the circular-shaped
coil. This makes it a good candidate for other MRI applications involving
coil array techniques. \bigskip{}

\end{abstract}

\section{Introduction}

Radio-frequency (RF) coils constitute the key hardware component for
the transmission and reception of the magnetic resonance signal. The
performance characteristics are a crucial element in the determination
of image quality as measured by the signal-to-noise ratio (\emph{SNR}),
signal homogeneity, and spatial resolution. The \emph{SNR} is the
widely-accepted parameter to measure coil performance since is independent
of the imaging parameters and the signal processing system. A great
deal of effort has been done to develop RF coils for different Magnetic
Resonance Imaging (MRI) and Magnetic Resonance Spectroscopy (MRS)
applications since the publication of the seminal papers by Hoult,
Richard and Lauterbur {[}1-2{]}. MRI scientists have proposed different
theoretical approaches to derive expressions for the \emph{SNR} models
involving low and high frequency approaches {[}3-9{]}. To find an
analytical solution even for simple coil geometries requires the use
of an intrinsicly-difficult mathematical frame. This has motivated
somehow to propose numerical solutions as an alternative method {[}10-11{]}.

The research work presented here was motivated by the renewed interest
in the development of RF surface coils caused by the introduction
of parallel imaging (PI) {[}12{]}. Parallel imaging requires of coil
arrays with high performance, however the sudy of the individual coil
performance plays an important role to achieve a high \emph{SNR} of
the coil array {[}13-14{]}.

A coil design based on the cavity magnetron tube {[}15-16{]} was introduced
by Rodriguez {[}17{]} , which showed a coil performance improvement
over the circular coil. An interesting historical fact, it is that
a magnetron tube was also used in the early NMR experiments to detect
the NMR signal by Purcell, Torrey and Pound back in 1945 {[}15{]}.
These experimental results obatined with the magnetron surface coil
stimulated used to calculate an \emph{SNR} formula to guide the further
development of this type of coils on a reliable manner. 

The objective of this work was to develop a theoretical model of the
signal-to-noise ratio for a the vane-type magnetron surface coil design
and named the slotted surface coil, see Fig. 1. \bigskip{}

\begin{center}
\includegraphics[scale=0.5]{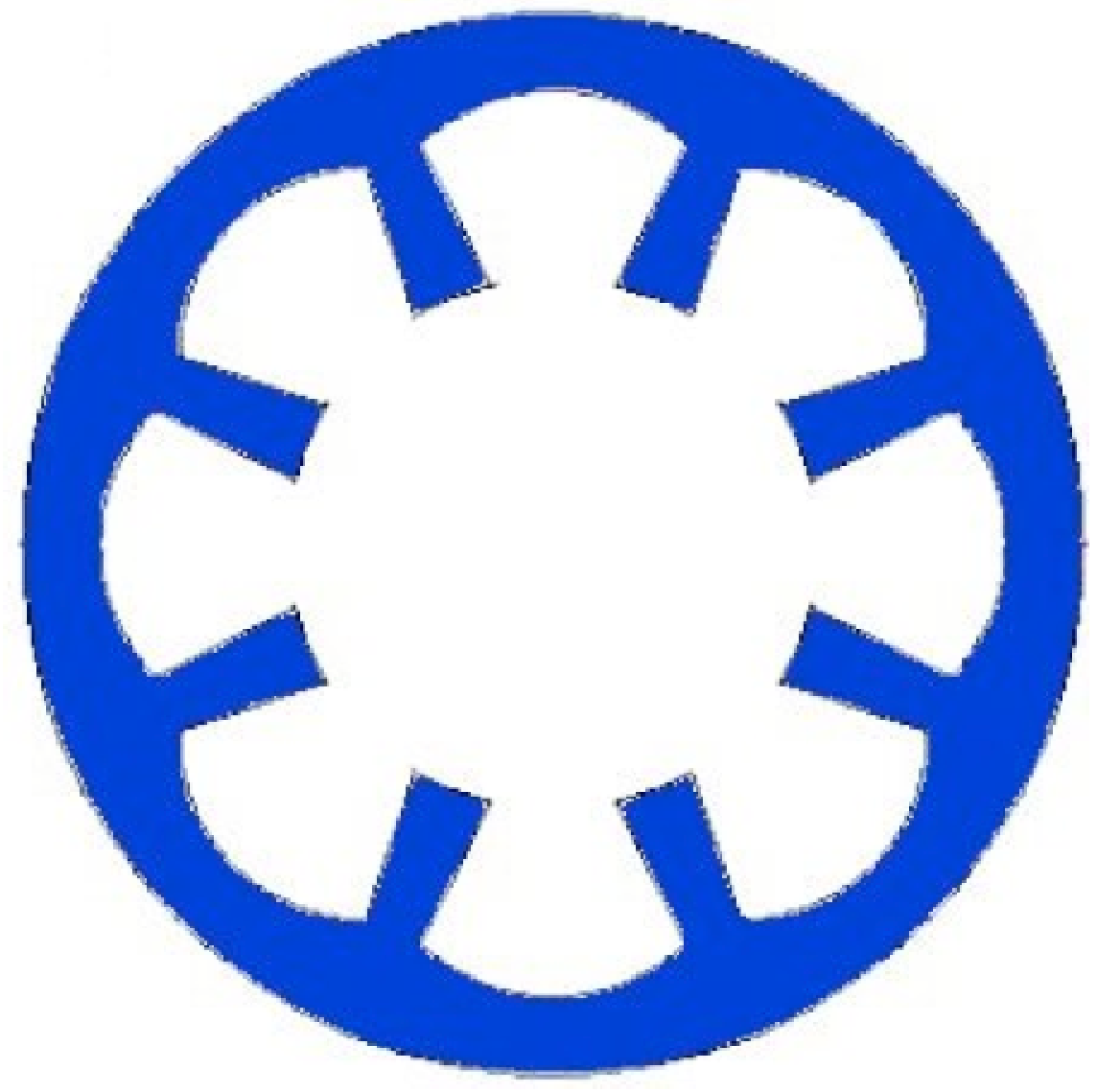}
\par\end{center}

\begin{figure}[h]
\selectlanguage{british}%
\centering{}\textbf{\small Figure 1. This illustration shows the vane
type configuration. }\selectlanguage{english}

\end{figure}

\bigskip{}
Then, a general \emph{SNR} expression was derived using the quasi-static
approach as a function of the number of vanes (slots). The formula
was used to investigate the performance of this coil design varying
the number of slots used for particular coil dimensions. It also served
to theoretically compare the coil performance against a circular-shaped
coil with similar dimensions for fare comparison.

\section{Signal-to-Noise Ratio (\emph{SNR})}

The signal-to-noise ratio determines the performance of a surface
coil, which is proportional to the ratio of the induced MR-signal
to the root-mean-square (rms) of the thermal noise voltage at the
coil terminals and the sample. According to Richard and Hoult {[}1{]},
the SNR can then be expressed 

\bigskip{}

\lyxaddress{\[
SNR(r)=\frac{\omega VMB_{1}(r)}{\sqrt{8kT\Delta f(P_{A}+P_{B})}}\qquad(1)\]
}

\bigskip{}

where $\omega$ is the Larmor frequency, \emph{V} is the sample volume\emph{,}
$B_{1}(r)$ is the magnetic field at position \emph{r} (Figure 2)\emph{,
k} is Boltzmann's constant, \emph{T} is the absolute temperature,
$\Delta f$ is the receiver bandwidth, $P_{A}$ and $P_{B}$ are the
power loss within the coil and the sample, respectively, when the
coil carries a given current \emph{I} or, more general, a given current
distribution \emph{J}. \emph{M} is the magnetisation perpendicular
to the static magnetic field and determined by the RF pulse sequence
{[}3{]}. To evaluate the \emph{SNR} of the slotted coil is necessary
to compute both the magnetic field and the power losses in Eq. (1).

\subsection{Calculation of Magnetic Field $B_{1}(r)$}

According to the quasi-static approach and Fig. 2, the magnetic field
$B_{1}(r)$ in Cartesian coordinates can be obtained from the Biot-Savart
law, \bigskip{}

\begin{center}
$B_{1}(r)={\displaystyle \frac{\mu_{0}}{4\pi}}{\displaystyle {\displaystyle {\displaystyle \int}}_{S'}}{\displaystyle \frac{J(r')\times(r-r')dS'}{\left|(r-r')\right|^{3}}}={\displaystyle \frac{\mu_{0}}{4\pi}}{\displaystyle {\displaystyle {\displaystyle \int}}_{S'}}{\displaystyle \frac{(J_{y}z)e_{x}+(-J_{x}z)e_{y}+(-J_{x}y'+J_{y}x')e_{z}\, dS'}{\left(x'^{2}+y'^{2}+z^{2}\right)^{3/2}}}\qquad(\textrm{2})$
\par\end{center}

\bigskip{}

where $J(r')=J_{x}e_{x}+J_{y}e_{y}$ is the current density and $r-r'=(-x')e_{x}+(-y')e_{y}+ze_{z}$.
Finally, \emph{$ $$ $}$r$ is the observation point and $r'$ is
the position of area element and $\mu_{0}$ is the permittivity.

\begin{center}
\bigskip{}
\includegraphics[scale=0.5]{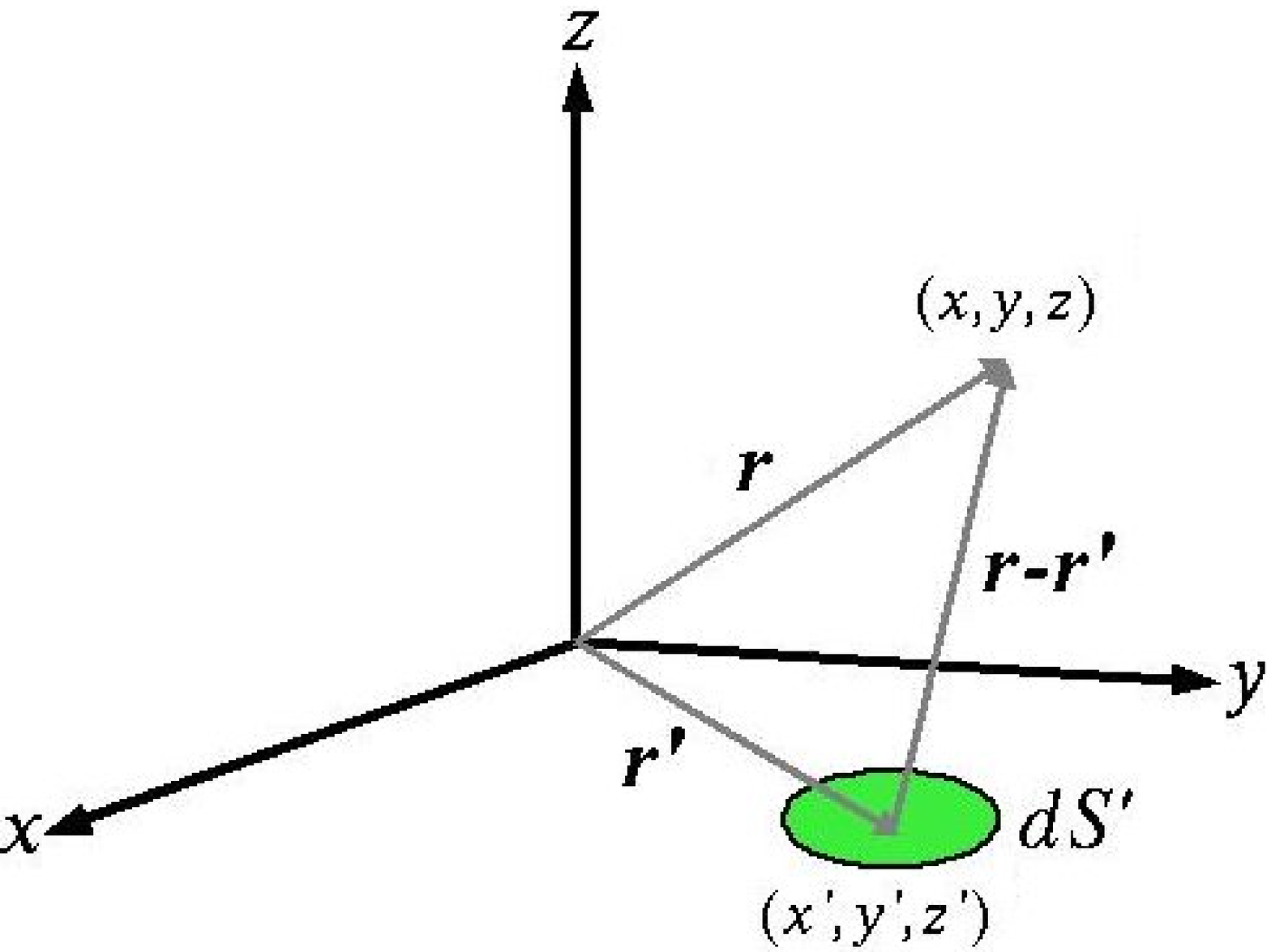}
\par\end{center}

\begin{figure}[h]
\selectlanguage{british}%
\centering{}\textbf{\small Figure 2. Schematic of geometric arrangement
for the computation of $B_{1}$ at observation point, $(x',y',z')$
created by an infinitesimal coil located at point, $r(x,y,z)$}\selectlanguage{english}

\end{figure}

\bigskip{}

From the coil geometry it is suggested to express Eq. (2) in cylindrical
coordinates \bigskip{}

\begin{center}
$\begin{array}{c}
\begin{array}{c}
B_{1}(z)={\displaystyle \frac{\mu_{0}}{4\pi}}\left[\begin{array}{c}
{\displaystyle {\displaystyle {\displaystyle {\displaystyle \int}}_{S'}\frac{J_{\phi}z\left(\cos(\phi)\cos(\phi')+\sin(\phi)\sin(\phi')\right)dS'}{\left(\rho'^{2}+z^{2}\right)^{3/2}}e_{\rho}}}\\
+{\displaystyle {\displaystyle {\displaystyle \int}}_{S'}\frac{J_{\phi}z\left(\cos(\phi)\sin(\phi')-\sin(\phi)\cos(\phi')\right)dS'}{\left(\rho'^{2}+z^{2}\right)^{3/2}}e_{\phi}}\\
+{\displaystyle {\displaystyle {\displaystyle \int}}_{S'}\frac{J_{\phi}\rho'dS'}{\left(\rho'^{2}+z^{2}\right)^{3/2}}e_{z}}\end{array}\right]\qquad(\textrm{3})\end{array}\end{array}$
\par\end{center}

\bigskip{}

If it is assumed that $dS'=\rho'd\rho'd\phi'$ the \emph{z}-component
of the magnetic field in Eq. (3) is \bigskip{}

\begin{center}
$\begin{array}{c}
\begin{array}{c}
B_{1}(z)={\displaystyle \frac{\mu_{0}}{4\pi}}{\displaystyle \int_{{\normalcolor \phi'}\epsilon\Re_{m}}}{\displaystyle \frac{J_{\phi}\rho'^{2}d\rho'd\phi'}{\left(\rho'^{2}+z^{2}\right)^{3/2}}}e_{z}\qquad(\textrm{4})\end{array}\end{array}$
\par\end{center}

\bigskip{}

wher $\Re_{m}$ is the integration area in Fig. 3.

\begin{center}
\includegraphics[scale=0.5]{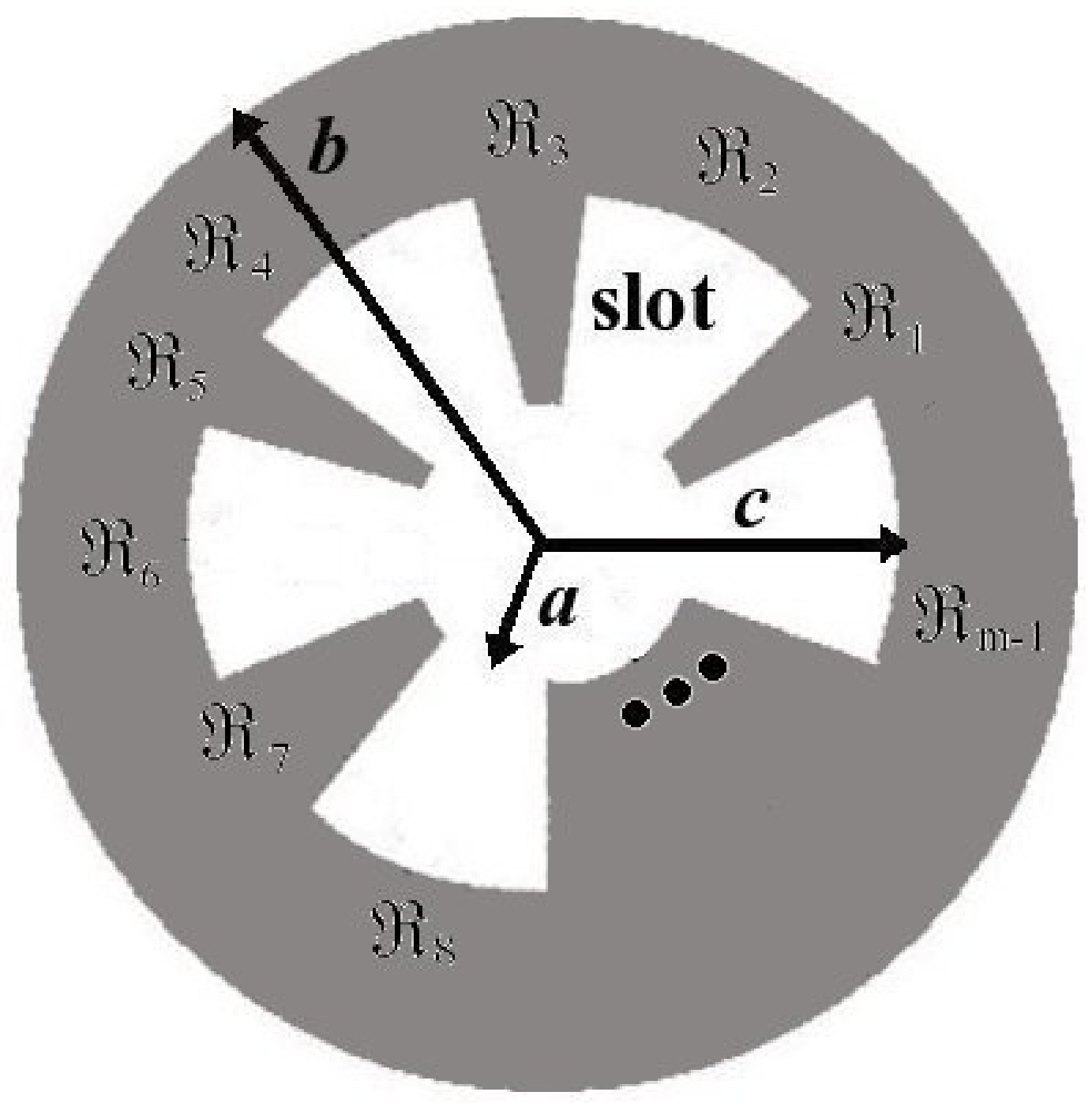}
\par\end{center}

\selectlanguage{british}%
\noindent \begin{center}
\textbf{\small Figure 3. Diagram of the slotted surface coil and design
parameters showing the area of integration. }
\par\end{center}{\small \par}

\selectlanguage{english}%
\bigskip{}

The linearly polarized magnetic field of the slotted surface coil
can be computed simirarly as in {[}19{]}, \bigskip{}

\begin{flushleft}
\[
\begin{array}{c}
\begin{array}{c}
{\displaystyle B_{1}(z)=\frac{\mu_{0}In}{2\pi}\left(\begin{array}{c}
\left.{\displaystyle \frac{\left(\frac{\pi}{n}-\varphi_{0}\right)}{(b-a)}\left(\frac{x}{\sqrt{x^{2}+z^{2}}}-\ln\left(-x+\sqrt{x^{2}+z^{2}}\right)\right)}\right|_{x=a}^{x=b}+\\
\\\qquad\qquad\left.\frac{\varphi_{0}}{(b-c)}\left(\frac{x}{\sqrt{x^{2}+z^{2}}}-\ln\left(-x+\sqrt{x^{2}+z^{2}}\right)\right)\right|_{x=c}^{x=b}\end{array}\right)e_{z}\quad(5)}\end{array}\end{array}\]

\par\end{flushleft}

\bigskip{}

where the parameters \emph{a}, \emph{c}, and $\varphi_{0}$ are defined
Fig. 3. The current circulating around the coil is \emph{I}, \emph{b}
is the coil radius\emph{, n} is the number of slots, and \emph{z}
is the depth point. A more general approach to investigate the slotted
coil performance should include a current distribution, $J_{\phi}$
which is considered uniform over the entire coil area.

\subsection{Calculation of the power losses $P_{A}$ and $P_{B}$}

The losses can be calculated by integration of the coil current distribution
over the area of the coil for $P_{A}$, and by integration of the
resulting electric field over the volume of the load for $P_{B}$.
The computation of the power losses was entirely based on the quasi-static
method introduced by Ocegueda and Rodriguez {[}19{]}. All calculations
were moved into Appendixes A and B.

The coil loss $P_{A}$ , was computed according to coil geometry in
Fig. 1c, and it can then be expressed, \bigskip{}

\begin{center}
$\begin{array}{c}
\begin{array}{c}
{\displaystyle {\displaystyle P_{A}=}{\displaystyle {\displaystyle \frac{I^{2}n}{\sigma_{A}\delta}\left(\frac{\left(b+a\right)\left(\frac{\pi}{n}-\varphi_{0}\right)}{b-a}+\frac{2\varphi_{0}\left(b+c\right)}{b-c}\right)}}{\displaystyle \qquad\qquad(6)}}\end{array}\end{array}$
\par\end{center}

\bigskip{}

where $\sigma_{A}$ is the coil conductivity. The sample loss $P_{B}$
is \bigskip{}

\begin{center}
$\begin{array}{c}
{\displaystyle {\displaystyle {\displaystyle P_{B}=}}}{\displaystyle \sigma_{B}\omega^{2}}{\displaystyle {\displaystyle \sum_{m=1}^{n}}}\begin{array}{c}
\left(\begin{array}{c}
{\displaystyle {\displaystyle \int_{\phi\epsilon\Re_{m}}\left(\int_{0}^{a}\left|A_{sc}\right|^{2}r^{2}dr{\displaystyle +{\displaystyle \int_{a}^{b}\left|A_{sc}\right|^{2}r^{2}{\displaystyle dr}}}+{\displaystyle {\displaystyle \int_{b}^{\infty}\left|A_{sc}\right|^{2}r^{2}\sin\left(\theta\right)dr}}\right)\sin\left(\theta\right)d\theta d\phi}}\\
\\+{\displaystyle {\displaystyle \int_{\phi\epsilon\Re_{m}}}}\left({\displaystyle {\displaystyle \int_{0}^{c}}}\left|A_{cc}\right|^{2}r^{2}dr{\displaystyle +{\displaystyle \int_{c}^{b}\left|A_{cc}\right|^{2}r^{2}{\displaystyle dr}}}+{\displaystyle {\displaystyle \int_{b}^{\infty}\left|A_{cc}\right|^{2}r^{2}\sin\left(\theta\right)dr}}\right)\sin\left(\theta\right)d\theta d\phi\end{array}\right)\end{array}\\
{\displaystyle \qquad\qquad\qquad\qquad\qquad\qquad\qquad\qquad\qquad\qquad\qquad\qquad\qquad\qquad\qquad\qquad\qquad\qquad(7)}\end{array}$
\par\end{center}

\bigskip{}

where $\sigma_{B}$ is the sample conductivity. $A_{sc}$ is the vector
potential due to the coil region between slots, and $A_{cc}$ is the
vector potential of the slot region.

Once the magnetic field and the power loss expressions were derived,
it is possible to calculate a general \emph{SNR} expression for the
slotted surface coil. Then, assuming that the sample resistance is
much greater than the coil resistance ($P_{B}\gg P_{A}$) {[}20{]},
Eq. (2) transforms into \bigskip{}

\begin{center}
$\begin{array}{c}
\begin{array}{c}
{\displaystyle SNR(z)=\frac{\omega VMB_{1}(z)}{\sqrt{8kT\Delta fP_{B}}}}\\
\\\qquad\qquad\qquad\qquad\qquad{\displaystyle =\frac{VM}{\sqrt{8kT\Delta f}}\frac{B_{1}(z)}{\sqrt{\sigma_{B}{\displaystyle {\displaystyle \int}}_{V}\left|A\right|^{2}dV}}\qquad(8})\end{array}\end{array}$
\par\end{center}

\bigskip{}

combining Eqs. (6) and (8) and replacing them in Eq. (1), Eq. (9)
then becomes \bigskip{}

\begin{center}
$SNR(z)=\frac{\omega VM\mu_{0}In}{2\pi\sqrt{8kT\Delta fP_{B}}}\left[\begin{array}{c}
\frac{1}{(b-a)}\left(\frac{a\sqrt{b^{2}+z^{2}}-b\sqrt{a^{2}+z^{2}}}{\sqrt{a^{2}+z^{2}}\sqrt{b^{2}+z^{2}}}+\ln\left(\frac{-a+\sqrt{a^{2}+z^{2}}}{-b+\sqrt{b^{2}+z^{2}}}\right)\right)\qquad\qquad\\
\\\qquad\qquad+\frac{1}{(b-c)}\left(\frac{c\sqrt{b^{2}+z^{2}}-b\sqrt{c^{2}+z^{2}}}{\sqrt{c^{2}+z^{2}}\sqrt{b^{2}+z^{2}}}+\ln\left(\frac{-c+\sqrt{c^{2}+z^{2}}}{-b+\sqrt{b^{2}+z^{2}}}\right)\right)\end{array}\right]\qquad(9)$
\par\end{center}

\bigskip{}

Particular \emph{SNR} formulae can derived for different coil dimensions
to study their behaviour as a function of depth from Eq. (9).

\section{Results and Discussion}

A general SNR expression of a slotted surface coil was derived using
the quasi static approach for \emph{n} vanes (Eq. 10). This formula
was used to theoretically compute \emph{SNR}-vs.-depth profiles to
study its behaviour as a function of the depth (\emph{z}) and various
coil design parameters. The profiles of Fig. 4 showed that there is
a clear improvement on the performance directly related to the number
of vanes for a particular set of parameters. \bigskip{}

\begin{center}
\includegraphics[scale=0.3]{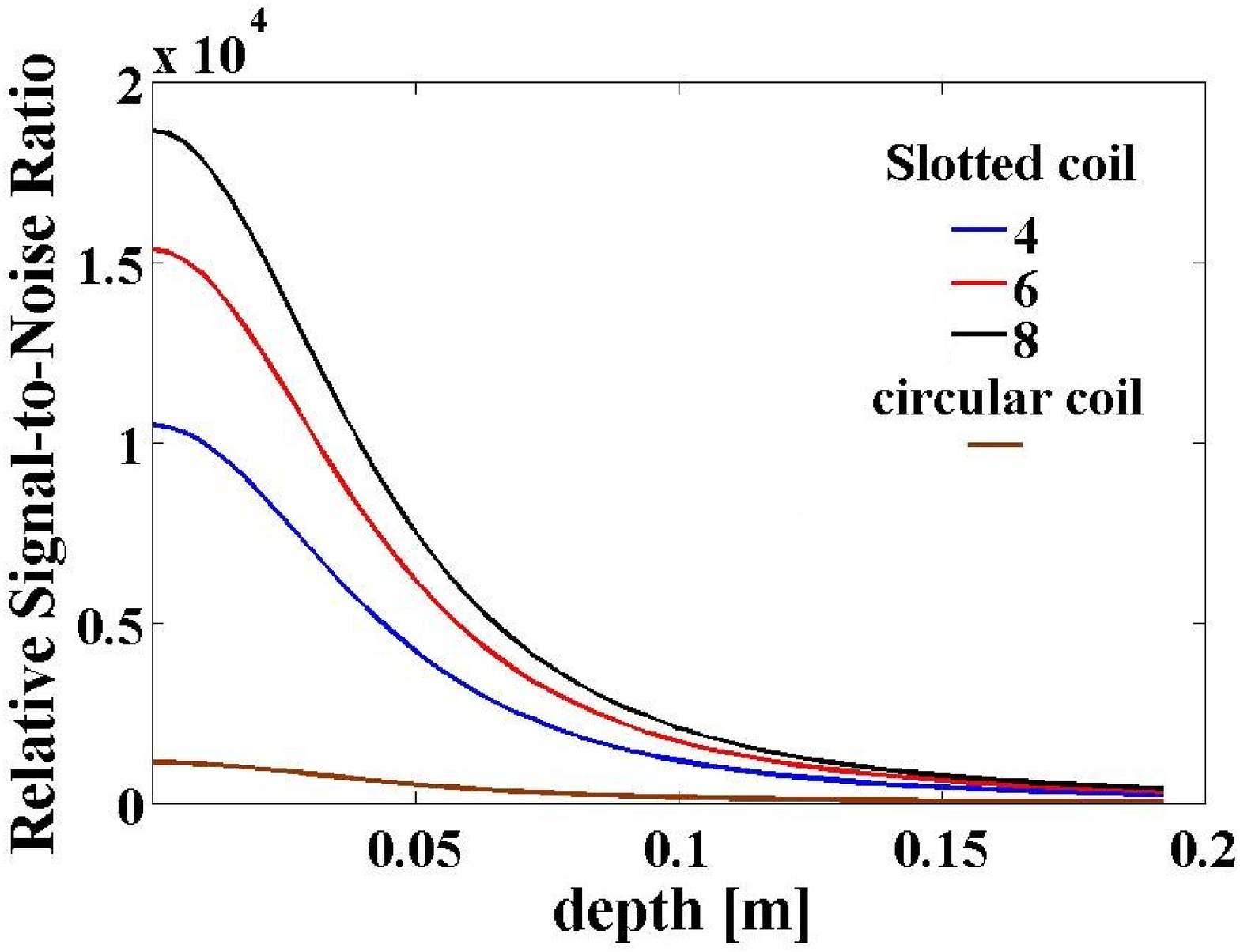}
\par\end{center}

\begin{figure}[h]
\selectlanguage{british}%
\centering{}\textbf{\small Figure 4. SNR profiles of the slotted surface
coil for 4, 6 and 8 vane-type slots and the following design parameters:
a = 2 cm, b (coil radius) = 6 cm, c= 4 cm. All SNR roll-offs were
computed using Eq. (10).}\foreignlanguage{english}{ }\selectlanguage{english}

\end{figure}
\bigskip{}

As \emph{n} increases a greater area is covered and the coil configuration
may be drastically affected by it. Extra care should be taken since
a specific size coil can not accommodate a high number of them. There
is a trade off between the number of vane-tyep slots and their size
that it is possible to accommodate in a particular coil size. To avoid
a decrement of the \emph{SNR}, coil dimensions should be ajusted to
fit more slots in a specific coil size. If a high number of slots
is required, a possible coil design candidate for this case may be
the coil configuration in Fig. 1b). This is an advantage over other
surface coil designs since it actually determines the way to theoretically
improve the \emph{SNR} by varying the number of slots and its size.

Additionally, to theoretically compare the slotted coil performance,
the \emph{SNR} profile of a circular-shaped coil was also calculated
using the quasi-static approach with similar dimensions for fare comparison.
There is no specific reason to choose 8 slots, other than following
the original design of the cavity magnetron tube, and that the previous
experimental results obtained at low frequency (64 MHz). From Fig.
4 the \emph{SNR} stills shows a substantially perfomance improvement
over the \emph{SNR} of a circular coil, for points located perpendicularly
further away from the coil plane up to the equivalent distance of
the total coil radius. The \emph{SNR} roll-offs theoretically outperformed
the circular-shaped coil. Additionally, \emph{SNR} plots as function
of the magnetic field intensity were computed and shown in Fig. 5. 

These plots exhibit a very similar pattern as that reported in the
literature for the low-field scheme. Therefore, the the \emph{SNR}
model of slotted coil and the correct selection of the parameters
can provide us with reliable guidelines to develop a surface coil
with an improved performance. The coil design presented in this work,
is only one of the many possible geometries that can be used.

The approach presented here has two important disadvantages, it is
only valid for low-field MRI since the quasi-static approach was employed
and that a uniform current distribution of the current densitiy was
used to derive the \emph{SNR} model. The classical electromagnetic
theory poses a great challenge if a more elaborate current distribution
is assumed for this coil design. As a first approximation to an intricately
difficult problem, an uniform distrubution was assumed despite the
fact that this is not accurate enough for a more realistic computation
of the magnetic field. 

\begin{center}
\bigskip{}
\includegraphics[scale=0.3]{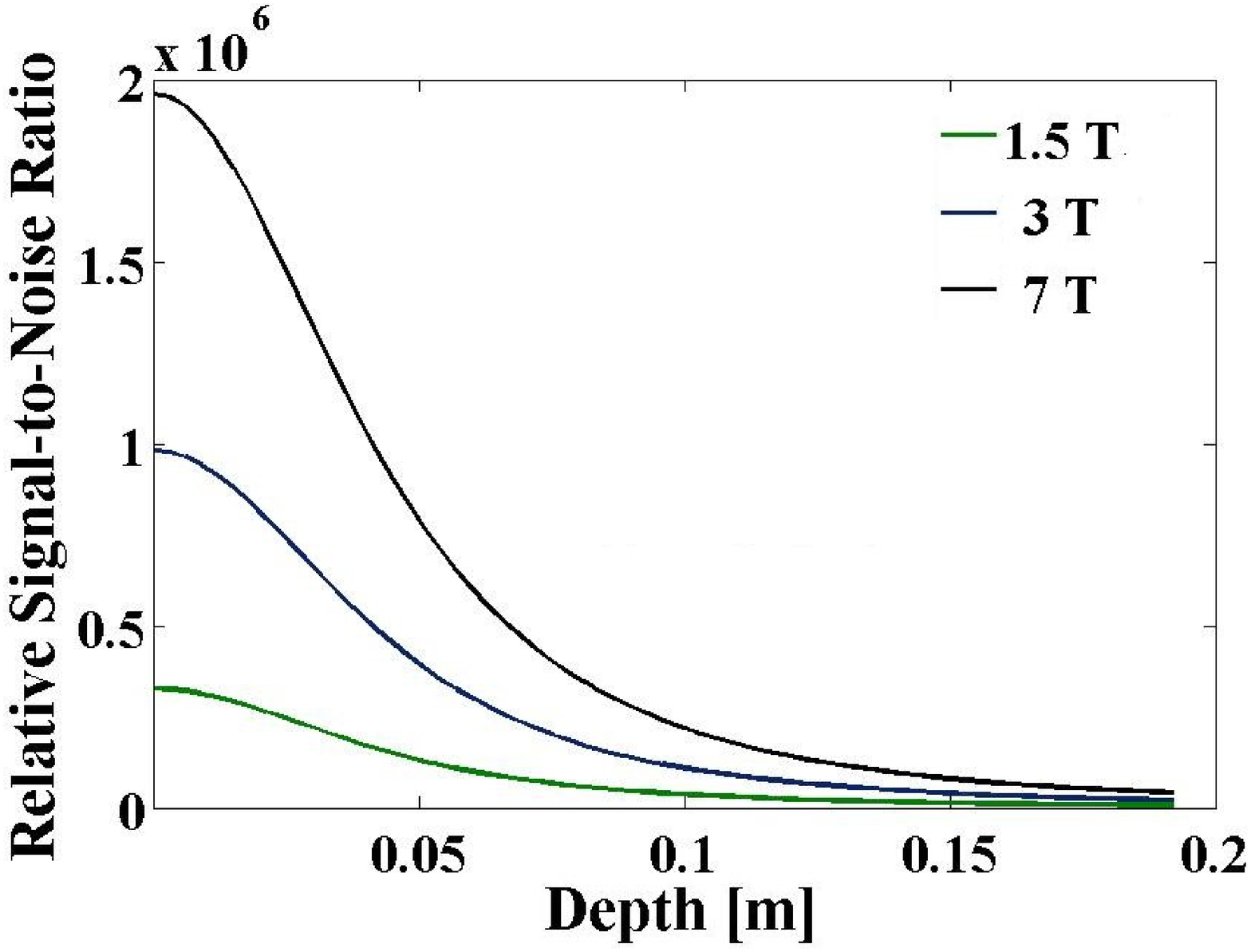}
\par\end{center}

\selectlanguage{british}%
\begin{center}
\textbf{\small Figure 5. Comparison of theoretically-acquired roll-offs
of relative SNR for the slotted surface coil with 8 slots at different
magnetic field intensities.}
\par\end{center}{\small \par}

\selectlanguage{english}%
\bigskip{}

Therefore, this low-field \emph{SNR} model of the slotted surface
coil should be viewed as a work in progress whose partial completion
has already shown a relatively good agreement with experimental data.
It remains then to investigate different approaches to derive other
\emph{SNR} models using a high-frequency approach with other coil
layouts. However limited this \emph{SNR} model, it can be in different
MRI applications such as parallel imaging and phased-array coils,
since it allows us to simulate their performance previously to actually
start developing coil array prototypes. 

This theoretical \emph{SNR} model allows us to simulate the coil performance
in advance of experiments thus saving a significant amount of effort
in designing these type of coils. This approach to surface receiver
coils can be used in MR imaging and MR spectroscopy.

\section{Conclusions}

The quasi static approach was used to calculate an \emph{SNR} expression
for a new coil design called the slotted surface coil. This general
\emph{SNR} expression permits to derive particular \emph{SNR} formulae
to study the performance of particular coil designs. From the theoretical
point of view, the slotted surface coil can produce an important improvement
compared with the single surface coil. These theoretical results showed
that the slotted surface coil design outperformed the standard circular-shaped
coil with similar dimensions. This approach is a good alternative
to other schemes used to develop MRI surface coils when dealing with
frequency values in the range of the quasi-static approach. This coil
design offers a new choice to existing surface coils for different
applications in both MRS and MRI.

\section*{Acknowledgments}

K. Ocegueda and S. E. Najera wish to thank the National Council of
Science and Technology of Mexico for Ph. D. scholarships. S. E. Najera
also acknowledges the Insitute of Science and Technology of Mexico
City for a posdoc stipend.

\subsection*{Appendix A}

\subsubsection*{Calculation of the power losses $P_{A}$}

To compute the power loss of the coil due to the Joule heating, $P_{A}$,
first calculate the average-time rate of work done by the electric
field on in the volume \emph{V} (homogenous conducting half-space,
$z>0$) as follows \bigskip{}

\begin{center}
${\displaystyle \frac{dP}{dV}=E\cdot J*=\frac{1}{\sigma}\left|J\right|^{2}=\sigma\left|E\right|^{2}\qquad\qquad(\textrm{A.1})}$
\par\end{center}

\bigskip{}

Eq. (A.1) was computed using the Poynting's theorem for harmonics
fields and Ohm's law. Therefore, the ratio of coil power loss per
volume unit is \bigskip{}

\begin{center}
${\displaystyle \frac{dP_{A}}{dV}=\frac{dP_{A}}{\delta dS'}=\frac{1}{\sigma_{A}}\left|J\right|^{2}\qquad\qquad(\textrm{A.2})}$
\par\end{center}

\bigskip{}

where $\sigma_{A}$ is the coil conductivity. From Eq. (A.2) the coil
power loss is \bigskip{}

\begin{center}
$\begin{array}{c}
\begin{array}{c}
\begin{array}{c}
{\displaystyle {\displaystyle P_{A}=\frac{\delta}{\sigma_{A}}\int_{S'}\left|J(r')\right|^{2}dS'=}}\\
\\={\displaystyle \frac{\delta}{\sigma_{A}}\sum_{m=1}^{n}\int_{{\normalcolor \phi'}\epsilon\Re_{m}}J_{\phi}^{2}\rho'd\rho'd\phi'=\frac{I^{2}n}{\sigma_{A}\delta}\left(\frac{\left(b+a\right)\left(\frac{\pi}{n}-\varphi_{0}\right)}{b-a}+\frac{2\varphi_{0}\left(b+c\right)}{b-c}\right)\qquad(}\textrm{A.3)}\end{array}\end{array}\end{array}$
\par\end{center}

\bigskip{}

Eq. (A.3) transforms into Eq. (3) which is the power loss $P_{A}$
for the \emph{n}-slot coil configuration of the slotted surface coil.

\bigskip{}

\subsubsection*{Appendix B}

\subsubsection*{Calculation of the power loss $P_{B}$}

Similarly as in the case of $P_{A}$, the ratio of power loss of $P_{B}$
per volume unit can be written as, \bigskip{}

\begin{center}
${\displaystyle \frac{dP_{B}}{dV}=\sigma_{B}\left|E\right|^{2}\qquad\left(\textrm{B.1}\right)}$
\par\end{center}

\bigskip{}

where $\sigma_{B}$ is the sample conductivity. To determine the electric
field \emph{E}, the Faraday-Lenz law can be used

\bigskip{}

\[
\bigtriangledown\times E=-{\displaystyle \frac{\partial B}{\partial t}={\displaystyle j\omega B}}\qquad({\textstyle \mathrm{B}}.2)\]

\bigskip{}

Eq. (B.2) holds true for time variation, $\exp(-jwt)$, and the field
\emph{B} can be rewritten $B=\bigtriangledown\times A$, so the electric
field \emph{E} transforms into

\bigskip{}

\[
E=-\bigtriangledown\Phi+jwA\qquad({\textstyle \mathrm{B}}.3)\]

\bigskip{}

where $\bigtriangledown\Phi$ is the scalar ptential and $E=j\omega A$
with \emph{A} being a vector potential. 

The sample power loss using the vector potential, \emph{A} in Eq.
(B.3) becomes \bigskip{}

\begin{center}
${\displaystyle {\displaystyle {\displaystyle P_{B}}=\sigma_{B}}{\displaystyle {\displaystyle {\displaystyle \int}}}_{V}{\displaystyle \left|E(r)\right|^{2}dV}={\displaystyle \sigma_{B}}{\displaystyle {\displaystyle {\displaystyle {\displaystyle \int}}}}_{V}{\displaystyle \left|j\omega A\right|^{2}dV=\sigma_{B}\omega^{2}}{\displaystyle {\displaystyle {\displaystyle \int}}_{V}}{\displaystyle \left|A\right|^{2}dV\qquad(\textrm{B.4})}}$
\par\end{center}

\bigskip{}

where $V=V_{1}+V_{2}$ and the semi-plane $z>0$. Eq. (B.4) is valid
if $\begin{array}{ccc}
\bigtriangledown\Phi=0\end{array}$ and rewriting \emph{A} in Cartesian coordinates \bigskip{}

\begin{center}
${\displaystyle A(r)=\frac{\mu_{0}}{4\pi}}{\displaystyle {\displaystyle \int}\frac{J(r')}{\left|r-r'\right|}}dS'={\displaystyle \frac{\mu_{0}}{4\pi}\int_{S'}}{\displaystyle \frac{J_{x}e_{x}+J_{y}e_{y}}{\left|\left(x-x'\right)^{2}+\left(y-y'\right)^{2}+z^{2}\right|^{1/2}}dS'}\qquad(\textrm{B.5})$
\par\end{center}

\bigskip{}

To facilitate the computation of the vector potential of Eq. (B.5),
the coil surface was split into two potentials corresponding to two
different areas: $A_{sc}$ (area surrounding the coil slots) and $A_{cc}$
(slot area). According to the principle of superposition, it is possible
to assume that the total coil area can be expressed as $A_{total}=A_{sc}+A_{cc}$,
where $A_{sc}$ corresponds to: $\Re_{1},\Re_{2},\Re_{3},\,\textrm{...\,}\Re_{n}$.

$A_{sc}$ can be rewritten in spherical coordinates as follows: \bigskip{}

\begin{center}
$\begin{array}{c}
\begin{array}{c}
{\scriptstyle {\displaystyle A_{sc}}={\displaystyle {\displaystyle {\scriptstyle \frac{J_{\phi}\mu_{0}}{4\pi}}\sum_{m=1}^{n}\left(\int_{{\normalcolor \phi'}\epsilon\Re_{m}}\int_{{\normalcolor r'}\epsilon\Re_{m}}{\scriptstyle \frac{-r'\sin\left(\phi'\right)}{\sqrt{r^{2}+r'^{2}-2r'r\cos\left(\gamma\right)}}}dr'd\phi'e_{x}+\int_{{\normalcolor \phi'}\epsilon\Re_{m}}\int_{{\normalcolor r'}\epsilon\Re_{m}}\frac{{\scriptstyle r'\cos\left(\phi'\right)}}{\sqrt{{\scriptstyle r^{2}+r'^{2}-2r'r\cos\left(\gamma\right)}}}dr'd\phi'e_{y}\right)}{\displaystyle \;(\textrm{B.6})}}}\end{array}\end{array}$
\par\end{center}

\bigskip{}

where $\cos\left(\gamma\right)=\sin\left(\theta\right)\sin\left(\phi-\phi'\right)$.
A similarly expression can be computed for $A_{cc}$. \bigskip{}

To solve integrals of Eq. (B.6) and the corresponding integral for
$A_{cc}$ , we can use the Legendre polynomials, this is a standard
procedure used to solve electrostatic problems. Then the denominator
of the integrand of (B.4) can be formulated in the frequently-used
Lengedre polynomials: \bigskip{}

\begin{center}
$\begin{array}{c}
{\displaystyle \frac{1}{\sqrt{r^{2}+(r')^{2}-2rr'\cos\left(\gamma\right)}}=\left\{ \begin{array}{c}
{\displaystyle \sum_{l=0}^{\infty}}\left(\frac{r^{l}}{r'^{l+1}}\right){\textstyle P_{l}\cos\left(\gamma\right)\qquad r<a}\\
\\{\displaystyle \sum_{l=0}^{\infty}}\left(\frac{r'^{l}}{r^{l+1}}\right){\textstyle P_{l}\cos\left(\gamma\right)\qquad r>b}\end{array}a<r'<b\right.\qquad(\textrm{B.7})}\end{array}$
\par\end{center}

\bigskip{}

\begin{center}
$\left.\begin{array}{cc}
{\displaystyle \frac{1}{\sqrt{r^{2}+(r')^{2}-2r'r\cos\left(\gamma\right)}}}={\displaystyle \sum_{l=0}^{\infty}}{\displaystyle \left(\frac{r'^{l}}{r'^{l+1}}\right)P_{l}\cos\left(\gamma\right)} & a<r'<r\\
\\+{\displaystyle \sum_{l=0}^{\infty}}{\displaystyle \left(\frac{r^{l}}{r'^{l+1}}\right)P_{l}\cos\left(\gamma\right)} & r<r'<b\end{array}\right\} a<r<b\qquad(\textrm{B.8})$
\par\end{center}

\bigskip{}

Therefore the vector potential of Eq. B.4 transforms \bigskip{}

\[
\begin{array}{c}
A_{sc}={\displaystyle \frac{{\scriptstyle J_{\phi}\mu_{0}}}{{\scriptstyle 4\pi}}{\displaystyle \sum_{l=0}^{\infty}{\displaystyle \sum_{m=1}^{n}}}\left[\begin{array}{c}
\left[{\displaystyle \int_{a}^{b}\left(\frac{r}{r'}\right)^{l}dr'+\int_{a}^{r}\left(\frac{r'}{r}\right)^{l+1}dr'+\int_{r}^{b}\left(\frac{r}{r'}\right)^{l}dr'+\int_{a}^{b}\left(\frac{r'}{r}\right)^{l+1}dr'}\right]\qquad\\
\\\left[{\displaystyle {\displaystyle -\int_{{\normalcolor \phi'}\epsilon\Re_{m}}\sin\left(\phi'\right)P_{l}\cos\left(\gamma\right)d\phi'e_{x}}+\int_{{\normalcolor \phi'}\epsilon\Re_{m}}\cos\left(\phi'\right)P_{l}\cos\left(\gamma\right)d\phi'e_{y}}\right]\end{array}\right]}\qquad(\textrm{B.9})\end{array}\]

\bigskip{}

In a complete similar fashion and expression for $A_{cc}$ can be
derived too.

\bigskip{}

Finally, the sampe noise $P_{B}$ becomes \bigskip{}

\[
\begin{array}{c}
{\displaystyle {\displaystyle {\displaystyle {\displaystyle P_{B}}=}{\displaystyle \sigma_{B}\omega^{2}}{\displaystyle {\displaystyle {\displaystyle \int}}_{V}}{\displaystyle \left|A\right|^{2}dV=}}{\displaystyle \sigma_{B}\omega^{2}}\left(\int_{V_{1}}\left|A_{sc}\right|^{2}dV_{1}+\int_{V_{2}}\left|A_{cc}\right|^{2}dV_{2}\right)}\qquad(\textrm{B.10})\end{array}\]

\bigskip{}

\end{document}